\documentclass[
  ,final            % use final for the camera ready runs
  ]
  {aipproc}

\include{epsf}
\layoutstyle{6x9}

\begin{document}

\newcommand{\lya}{Ly$\alpha$}
\newcommand{\kms}{\;{\rm km}\;{\rm s}^{-1}}
\newcommand{\invkms}{\;({\rm km}\;{\rm s}^{-1})^{-1}}
\newcommand{\hmpc}{h^{-1}\;{\rm Mpc}}
\newcommand{\meand}{\langle D \rangle}
\newcommand{\apj}{ApJ}
\newcommand{\aj}{AJ}
\newcommand{\mnras}{MNRAS}

\title{The Lyman-$\alpha$ Forest As A Cosmological Tool}

\author{David H. Weinberg}{
  address={The Ohio State University, Dept. of Astronomy, Columbus, OH 43210}
}

\author{Romeel Dav\'e}{
  address={University of Arizona, Dept. of Astronomy, Tucson, AZ 85721}
}
\author{Neal Katz}{
  address={University of Massachusetts, Dept. of Physics and Astronomy, Amherst, MA, 91003}
}
\author{Juna A. Kollmeier}{
  address={Ohio State University, Dept. of Astronomy, Columbus, OH
43210} }

\begin{abstract}
We review recent developments in the theory of the \lya\ forest and their
implications for the role of the forest as a test of cosmological models.
Simulations predict a relatively tight correlation between the local \lya\ 
optical depth and the local gas or dark matter density. Statistical properties
of the transmitted flux
can constrain the amplitude and shape of the matter power spectrum at high
redshift, test the assumption of Gaussian initial conditions, and probe the
evolution of dark energy by measuring the Hubble parameter $H(z)$. Simulations
predict increased \lya\ absorption in the vicinity of galaxies, but observations
show a \lya\ deficit within $\Delta_r \sim 0.5\hmpc$ (comoving). We investigate
idealized models of ``winds'' and find that they must eliminate neutral hydrogen
out to comoving radii $\sim 1.5\hmpc$ to marginally explain the data. Winds of
this magnitude suppress the flux power spectrum by $\sim 0.1$ dex but have 
little effect on the distribution function or threshold crossing frequency.
In light of the stringent demands on winds, we consider the alternative 
possibility that extended \lya\ emission from target galaxies replaces
absorbed flux, but we conclude that this explanation is unlikely. Taking full
advantage of the data coming from large telescopes and from
the Sloan Digital Sky
Survey will require more complete understanding of the galaxy proximity effect,
careful attention to continuum determination, and more accurate numerical 
predictions, with the goal of reaching $5-10\%$ precision on key
cosmological quantities.
\end{abstract}

\maketitle

%%%%%%%%%%%%%%%%%%%%%%%%%%%%%%%%%%%%%%%%%%%%
%% MAINMATTER
%%%%%%%%%%%%%%%%%%%%%%%%%%%%%%%%%%%%%%%%%%%%

\section{Physics of the Forest}

The 1990s saw four epochal advances in our understanding of the
\lya\ forest.  Spectra of quasar pairs showed coherence over scales
of a hundred kpc and more, implying large sizes and thus low densities
for the absorbing structures \cite{bechtold94,dinshaw94,dinshaw95,crotts98}.  
Keck HIRES spectra of
unprecedented resolution and signal-to-noise demonstrated the
ubiquity of weakly fluctuating \lya\ absorption in the high redshift
universe \cite{hu95}, and they revealed the presence
of metal lines associated with low column density hydrogen absorbers
\cite{songaila96,ellison99}.  
Finally, and most directly relevant to this review, a
combination of numerical simulations and related analytic models led to
a compelling new physical picture of \lya\ forest absorption 
\cite{bi93,cen94,petitjean95,zhang95,hernquist96,bi97,hui97,theuns98}.

\begin{figure}
\centerline{
\epsfxsize=5.0truein
\epsfbox[40 10 565 740]{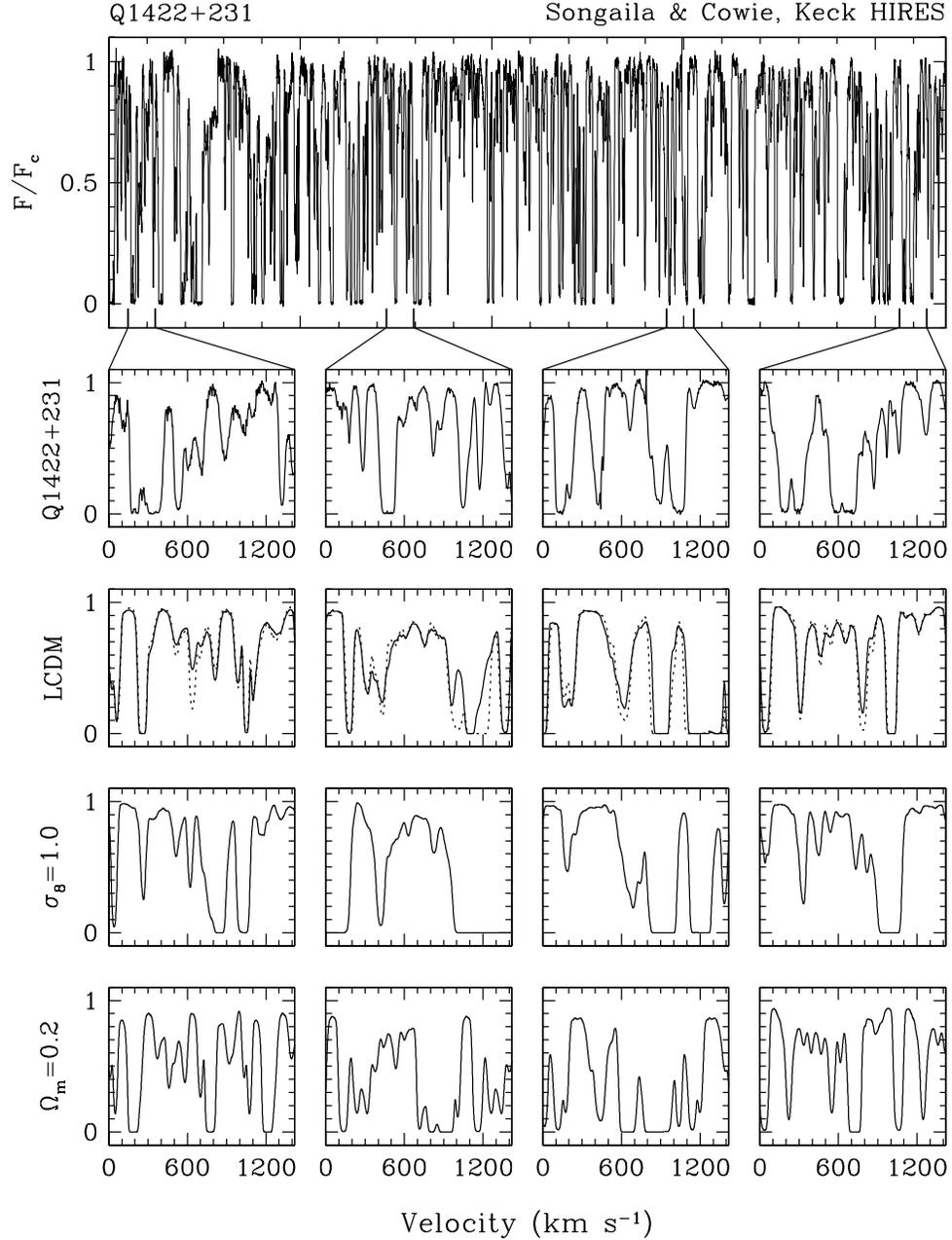}
}
\caption{
Simulating the \lya\ forest.  The top panel shows a continuum normalized,
Keck HIRES spectrum of the \lya\ forest region of the quasar Q1422+231
($z=3.62$), from \cite{songaila96}.
The next row of panels shows blowups of four regions 
$1422\kms$ in length.  The third row shows simulated spectra extracted
along four random lines of sight at $z=3$ from an SPH simulation (solid) and a
PM simulation (dotted) of the $\Lambda$CDM model, with $\Omega_m=0.4$, $h=0.65$,
and $\sigma_8=0.8$ (at $z=0$).  The fourth and fifth rows show PM results
for different cosmological parameter values: $\sigma_8=1.0$ and 
$\Omega_m=0.2$, respectively.  All simulated spectra are $1422\kms$
in length.
}
\label{fig:spec}
\end{figure}

The basic numerical result is simple to summarize: 
given a cosmological scenario motivated by independent observations,
3-d simulations that incorporate gravity, gas dynamics, and photoionization 
by the UV background produce something very much like the observed \lya\
forest, an outcome that requires no {\it ad hoc} adjustments
to the model.
The top three rows of Figure~\ref{fig:spec} illustrate this 
point, showing, respectively, the observed \lya\ forest of the $z=3.62$ quasar
Q1422+231, expanded views of four selected regions
of this spectrum, and simulated spectra of the same length along four randomly 
selected lines of sight through a cosmological simulation.
The simulation uses smoothed particle hydrodynamics (SPH) with
$128^3$ dark matter particles and $128^3$ gas particles in a periodic
cube of comoving size $11.111\hmpc$ ($1422\kms$ at $z=3$).
It assumes a $\Lambda$CDM model
(inflationary cold dark matter with a cosmological constant), 
with $\Omega_m=0.4$, $\Omega_\Lambda=0.6$, $h=0.65$,
$\Omega_b=0.02h^{-2}=0.0473$, inflationary spectral tilt $n=0.95$,
and a power spectrum normalization that corresponds to $\sigma_8=0.8$
at $z=0$.  Spectra are extracted 
from the $z=3$ simulation output using the methods of \cite{hernquist96}.  

Studies of the \lya\ forest have traditionally focused on ``lines'' identified
by a decomposition procedure and the ``clouds'' or ``absorbers'' that
produce them.  Hydrodynamic simulations show that typical
marginally saturated absorption features at $z\sim 3$ arise in
filamentary structures, which are analogous to (but smaller in scale than)
today's galaxy superclusters.  However, in the simulations there is no
sharp distinction between the ``lines'' and the ``background,''
and one can also characterize the \lya\ forest as ``Gunn-Peterson'' 
\cite{gunn65} absorption produced by a smoothly fluctuating
intergalactic medium \cite{hernquist96,croft97,rauch97,weinberg98}.

What makes the fluctuating IGM perspective especially powerful is
the tight relation between the density and temperature of
low density cosmic gas, $T \approx T_0(\rho/\bar{\rho})^\alpha$ with 
$\alpha\approx 0.6$, which emerges from the balance between photoionization 
heating and adiabatic cooling \cite{huig97,weinberg97}.
The \lya\ optical depth of photoionized gas 
is $\tau \propto n_{\rm HI} \propto \rho^2 T^{-0.7}\Gamma^{-1}$,
where $T^{-0.7}$ accounts for the temperature dependence of
the recombination rate and $\Gamma$ is the rate at which neutral
atoms are ionized by the cosmic UV background.
The temperature-density relation then leads to 
the ``Fluctuating Gunn-Peterson Approximation'' (FGPA), 
$F \equiv \exp(-\tau) = \exp[-A(\rho/\bar{\rho})^\beta]$, which 
relates the continuum normalized flux $F$ to the local gas overdensity
\cite{croft97,croft98,rauch97,weinberg98}.  
The index $\beta$ lies in the range $1.6-2$ depending on the thermal 
history of the gas, and the proportionality constant $A$ is itself 
proportional to $\Omega_b^2 h^3 T_0^{-0.7} \Gamma^{-1}$.
Thus, one can think of a \lya\ forest spectrum
as providing a 1-dimensional, non-linear map of the gas overdensity 
$\rho/\bar{\rho}$ along the line of sight.  The map is smoothed by
thermal broadening and distorted by peculiar velocities, but these
effects are small enough to leave a tight correlation between
\lya\ optical depth and local gas density even in redshift space
\cite{croft97,weinberg99}.
Furthermore, pressure gradients in the
diffuse, photoionized IGM are usually weak compared to gravitational forces,
so the gas overdensity traces the dark matter overdensity fairly well.

The FGPA provides a valuable way of thinking about the information content
of the \lya\ forest, and it can be a useful calculational tool if one has
a way to produce realizations of cosmic density and velocity fields.
The analytic models of the \lya\ forest by 
\cite{bi93,bi95,bi97,hui97} essentially 
combine the FGPA with a log-normal or Zel'dovich approximation
model for creating non-linear density
and velocity fields.  Alternatively, one can run an inexpensive, 
gravity-only N-body simulation and assume that the diffuse gas traces
the dark matter (or add an approximate treatment of gas pressure 
\cite{gnedin98,zaldarriaga01,viel02}).  The dotted lines in the third row of 
Figure~\ref{fig:spec} show spectra extracted from a particle-mesh (PM) 
N-body simulation, run from the same initial conditions as the SPH simulation, 
illustrating both the effectiveness and the limitations
of this approach.  The two spectra trace each other over most of their
length, with the largest breakdowns occurring in regions where 
shock heating has pushed the gas above the $T\propto \rho^\alpha$
temperature-density relation, which makes the optical depth in the 
SPH simulation lower than the FGPA predicts.

The N-body+FGPA technique offers a convenient way to investigate 
the response of the \lya\ forest to changes in cosmological parameters.
Comparing the third and fourth rows of Figure~\ref{fig:spec}
illustrates the effect of increasing the matter fluctuation amplitude
by 20\%, to $\sigma_8(z=0)=1.0$, with the same Fourier phases in the 
initial conditions.
The increased clustering of the high amplitude model is evident in
the greater incidence of saturated absorption and in the merging of
multiple small scale features into single larger scale features.
The bottom row shows spectra from a model with $\sigma_8=0.8$  
and the same linear power spectrum but a lower matter density,
$\Omega_m=0.2$.  The reduction in $\Omega_m$
lowers the value of the Hubble parameter $H(z)$ at $z=3$, and as a
result features are more densely packed in redshift space, producing
``choppier'' spectra.  In fact, our $11.111\hmpc$ simulation cube is
only $1024\kms$ in length for $\Omega_m=0.2$, and we have ``padded''
each spectrum to $1422\kms$ by replicating the first $400\kms$.
Reducing $\Omega_m$ also raises the fluctuation amplitude at $z=3$
(since $\sigma_8$ is held fixed at $z=0$ and there is less late time growth
for lower $\Omega_m$), but this effect is less important than the
effect on the Hubble parameter.

The relative simplicity of the underlying physics and the existence
of superb data at redshifts that are only sparsely probed by other
observables make the \lya\ forest a potentially powerful tool
for testing cosmological models.  Given independent estimates of 
$\Gamma$ (e.g., from the quasar luminosity function or the proximity 
effect), one can use the mean opacity of the forest to obtain a lower 
limit to the cosmic baryon density \cite{rauch97,weinberg97b}.
However, the observational constraints on $\Gamma$ are loose,
so for higher precision applications one must treat $\Gamma$ as a free 
parameter and adjust it to match a single observable, usually taken to
be the mean opacity.  This normalization also absorbs uncertainties in
$\Omega_b$, $T_0$, and $h$.  
We have followed this practice for each of the simulations in
Figure~\ref{fig:spec} and will adopt it in all of our subsequent analyses.
After normalizing to the mean opacity, the variation about the mean ---
i.e., the structure of the \lya\ forest ---
is a prediction of the cosmological model, driven primarily by 
the structure in the underlying dark matter distribution.
Ideally, one would like to calculate predictions for each cosmology using high
resolution hydrodynamic simulations, but for some purposes other
numerical or analytic approximations may be accurate enough.
The adequacy of such approximations must be tested on a case-by-case
basis, and the improving quality of the observational data places
ever more stringent demands on the accuracy of the theoretical
calculations.

\section{Continuous Flux Statistics}

Because of the tight predicted correlation between the 
local \lya\ optical depth
and the local gas density, the most natural statistics
for characterizing the \lya\ forest are those that treat each
quasar spectrum as a continuous 1-dimensional field, rather than
a collection of discrete ``lines.''  We have reviewed this
approach elsewhere \cite{weinberg98,weinberg99}, and here
we provide a brief recap with different examples and emphasis.

\begin{figure}
\centerline{
\epsfxsize=3.0truein
\epsfbox[90 415 460 720]{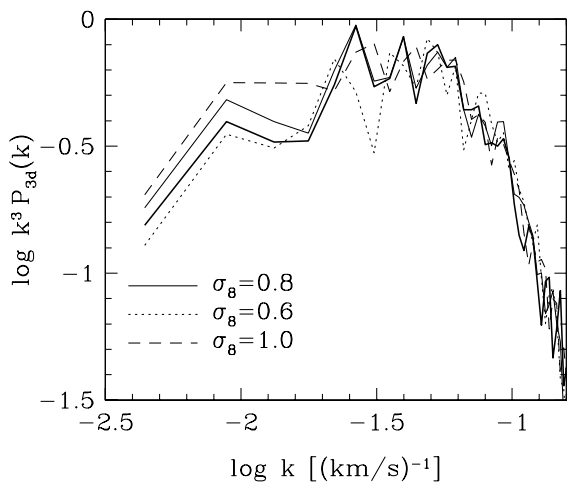}
}
\caption{
The flux power spectrum.  Heavy and light solid curves show the 3-d
flux power spectrum from, respectively, the SPH simulation and a PM 
simulation with the same initial conditions.  Dotted and dashed 
curves show results with lower and higher matter fluctuation amplitudes,
corresponding to $\sigma_8=0.6$ and 1.0 at $z=0$.
All results are from a single $11.111\hmpc$ cube realized with the
same Fourier phases.
}
\label{fig:fluxpk}
\end{figure}

Figure~\ref{fig:fluxpk} shows the power spectrum of continuum normalized
flux, converted from 1-d to 3-d as described by \cite{croft98}, and
multiplied by $k^3$ to yield the variance per ln$k$.  The curves are
noisy because they are based on a single $11.111\hmpc$ comoving cube,
but we use the same phases in each simulation, so the relative behavior
should be minimally affected by noise.  At scales $k > 0.03\invkms$,
the power spectra turn over because of the combined effects of non-linearity,
peculiar velocities, and thermal broadening.  The predictions in this
regime are also affected by the finite resolution of the simulations.
At larger scales, the SPH and N-body+FGPA methods give similar but
not identical results for the same cosmological model, and the 
amplitude of the flux power spectrum increases with the amplitude
of the matter power spectrum, as expected based on the FGPA and on
Figure~\ref{fig:spec}.  

On large scales, the shape of the 3-d flux power spectrum is similar to that
of the linear matter power spectrum $P_m(k)$ \cite{croft98,croft02}.
The close connection between the shape and amplitude of the flux 
power spectrum and the shape and amplitude of $P_m(k)$
is the basis of Croft et al.'s \cite{croft98} 
method for recovering $P_m(k)$ from \lya\ forest data.
Applying this method to a sample of 30 Keck HIRES spectra and 23 Keck
LRIS spectra yields a matter power spectrum in remarkably (or, 
perhaps, disappointingly)
good agreement with the ``concordance'' $\Lambda$CDM model favored
by CMB, supernova, weak lensing, and low-$z$ large scale structure data
(\cite{croft02}; see 
\cite{gnedin02} for an independent analysis of the same flux power spectrum
and \cite{tegmark02} for an independent comparison to other cosmological 
constraints).  McDonald et al.\ \cite{mcdonald00} reach similar conclusions 
from a ``forward'' comparison of hydrodynamic simulation predictions
to the flux power spectrum measured from eight HIRES
spectra.  The \lya\ forest power spectrum tests the $\Lambda$CDM model
in a previously unexplored regime of redshift and lengthscale.
It confirms one of the scenario's key predictions, a linear power spectrum
that bends from the primeval $k^n$ towards $k^{n-4}$ on small scales.
The implied constraints on cosmological parameter combinations complement
those from the CMB and other data.  The Sloan Digital Sky Survey (SDSS)
has already obtained moderate resolution spectra of several thousand
high redshift quasars, and these will soon provide measurements
of the flux power spectrum with much greater precision on large scales.

\begin{figure}
\centerline{
\epsfxsize=4.5truein
\epsfbox[60 470 555 720]{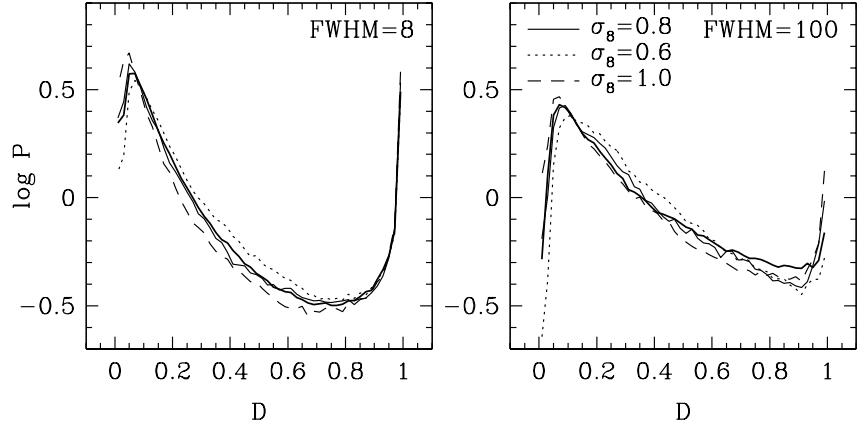}
}
\caption{
PDF of the flux decrement
$D=1-e^{-\tau}$, measured from simulated spectra smoothed
with a Gaussian of FWHM=$8\kms$ (left) or $100\kms$ (right).
Line types as in Fig.~\ref{fig:fluxpk}.
}
\label{fig:fdf}
\end{figure}

Figure~\ref{fig:fdf} shows a different statistic, the probability
distribution function (PDF) of the flux decrement
$D = 1-F = 1-e^{-\tau}$, for the same set of simulations.
The measurements in the left hand panel are from simulated spectra
smoothed with a Gaussian of ${\rm FWHM}=8\kms$, comparable to
the resolution of Keck HIRES or VLT UVES spectra.  The SPH and
N-body+FGPA predictions agree well for $\sigma_8=0.8$.
Models with higher fluctuation amplitude have a broader distribution
of densities $\rho$ and a correspondingly broader distribution 
of flux decrements $D$, with more saturated and low-opacity pixels
and fewer pixels of intermediate opacity.  The differences in
the predicted PDFs for $D\sim 0.3-0.7$ are $\Delta{\rm log}P\sim 0.1$ ---
not enormous, but readily measurable at high statistical significance with
reasonable observational samples.  The fraction of saturated
pixels (defined here by $D>0.96$) increases from 7.2\% to 8.3\% to 9.1\%
as $\sigma_8$ goes from 0.6 to 0.8 to 1.0.
There are larger differences in the number of nearly transparent
pixels, but continuum fitting uncertainties make it difficult
to measure the PDF accurately near $D=0$.
One can investigate the scale dependence of matter clustering
by smoothing the spectra (or by observing them at lower spectral resolution),
analogous to studying galaxy counts in cells of increasing size.
Figure~\ref{fig:fdf} (right) shows results for $100\kms$ smoothing.
The predicted PDFs are narrower, since smoothing drives pixel
values towards the mean, but the dependence
on $\sigma_8$ is similar.  The saturated pixel fraction doubles,
from 1.9\% to 3.8\%, as $\sigma_8$ rises from 0.6 to 1.

For cosmological models with Gaussian initial conditions, the 
shape of the flux PDF depends mainly on the amplitude of mass fluctuations,
once one has chosen $\Gamma$ to match the mean opacity,
\cite{cen97,weinberg99}.  The effective physical scale of this
amplitude measurement is determined by
the spectral smoothing or, for spectra that fully resolve the observed
absorption features,
by a combination of thermal broadening and gas pressure effects.
Models with non-Gaussian initial conditions predict significantly
different flux PDFs \cite{weinberg99}.  
Observational tests to date show good agreement with 
models that have Gaussian 
initial conditions and a $P_m(k)$ amplitude compatible with
$\Lambda$CDM predictions \cite{rauch97,weinberg99,mcdonald00}.
The uncertainties in these tests are comparable to the model
differences in Figure~\ref{fig:fdf}.  
Reducing them
requires careful attention to the effects of continuum fitting and 
noise and to the accuracy of the theoretical predictions, but it
should be possible to obtain tight constraints on the normalization
of the matter power spectrum, and some information on its shape
by studying different smoothing scales.  Consistency between results
from the flux power spectrum and the flux PDF can be a sensitive
diagnostic for primordial non-Gaussianity.

\begin{figure}
\centerline{
\epsfxsize=4.5truein
\epsfbox[60 470 555 720]{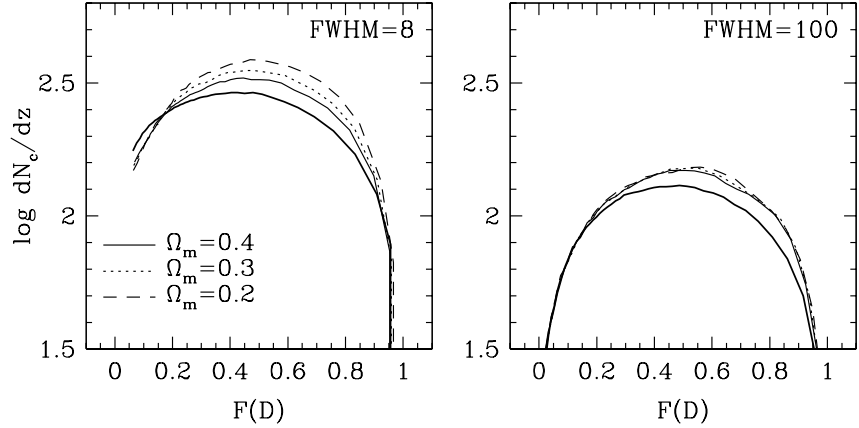}
}
\caption{
Threshold crossing frequency in spectra with FWHM=$8\kms$ (left)
and $100\kms$ (right).  Curves show the number of times per unit
redshift that the spectrum crosses a threshold of flux decrement $D$
as a function of the filling factor, the fraction $F(D)$ of
pixels with flux decrement less than $D$.  Heavy and light solid curves
show the SPH and PM results, respectively, for $\Omega_m=0.4$.
Dotted and dashed lines show PM results for $\Omega_m=0.3$ and 0.2,
respectively.
}
\label{fig:dcross}
\end{figure}

Figure~\ref{fig:dcross} shows the threshold crossing frequency, the
number of times per unit redshift that the absorption spectrum crosses
a decrement threshold $D$.  Following \cite{miralda96}, we plot
$dN_c/dz$ as a function of filling factor, the fraction $F(D)$ of pixels
with flux decrement less than $D$, which cleanly separates the information 
in $dN_c/dz$ from the information in the PDF and makes the model predictions
nearly independent of the photoionization rate $\Gamma$.  
As one would expect from Figure~\ref{fig:spec}, the threshold crossing
frequency increases as $\Omega_m$ decreases because a lower Hubble
parameter $H(z)$ ``squeezes'' redshift 
separations relative to comoving distances.
The threshold crossing frequency drops slightly as the 
amplitude of $P_m(k)$ increases and gravitationally driven merging
``smooths'' structure \cite{weinberg99}, but the effect is small
over the range $\sigma_8=0.6-1$.  The crossing frequency also drops
for redder matter power spectra, since these also lead to smoother
structure \cite{weinberg99}.  For high spectral resolution, 
$dN_c/dz$ also depends on the gas temperature, which determines
the level of thermal broadening.

The great promise of the threshold crossing statistic is its potential
for constraining the Hubble parameter at high redshift.  In spatially
flat models with a cosmological constant, the ratio $H(z)/H_0$ is
determined by $\Omega_m$.  Alternatively, if $\Omega_m$ is known
independently, the ratio $H(z)/H_0$ can constrain the equation of
state of dark energy \cite{kujat02,viel02c}.  Unfortunately, $dN_c/dz$ also
depends on other cosmological and IGM parameters, and it is numerically
difficult to predict with high accuracy even when the model is fully
specified.  The difference between the PM and SPH predictions in 
Figure~\ref{fig:dcross} is comparable to the $\Omega_m$ effects themselves,
and the SPH result is still affected by the finite numerical resolution.
On the observational side, accurate measurement of $dN_c/dz$ requires
high signal-to-noise spectra.  Furthermore, while moderate resolution
spectra can provide useful diagnostics for the shape and amplitude of
the matter power spectrum \cite{weinberg99}, the $H(z)$ application
demands high spectral resolution, since smoothing over a scale that
is fixed in redshift units erases the sensitivity to $H(z)$ 
(see Fig.~\ref{fig:dcross}, right).  Exploiting $dN_c/dz$ as a probe
of dark energy thus represents a theoretical and observational challenge.

\section{The Galaxy Proximity Effect}

One can also use the observable correlations between Lyman Break Galaxies
(LBGs) and the \lya\ forest to study the environments of high redshift
galaxies.  Several groups have investigated this issue theoretically
using simulations \cite{mcdonald02,croft02b,kollmeier02,bruscoli02},
and Adelberger et al.\ have carried out an observational study using
an LBG survey in fields probed by seven quasar lines of sight
(\cite{adelberger02}, hereafter ASSP).
On large scales --- cubes of comoving size $\sim 13\hmpc$ --- ASSP
find a clear correlation of galaxy overdensity with \lya\ flux
decrement, the expected signature of galaxy formation in overdense
environments \cite{mcdonald02,kollmeier02}.  However, the observed
\lya\ decrement {\it decreases} within $\Delta_r\approx 1\hmpc$ of
LBGs (comoving, redshift-space separation), where the simulations
predict that absorption should be strongest \cite{croft02b,kollmeier02,
bruscoli02}. 

\begin{figure}
\centerline{
\epsfxsize=4.5truein
\epsfbox[60 270 455 720]{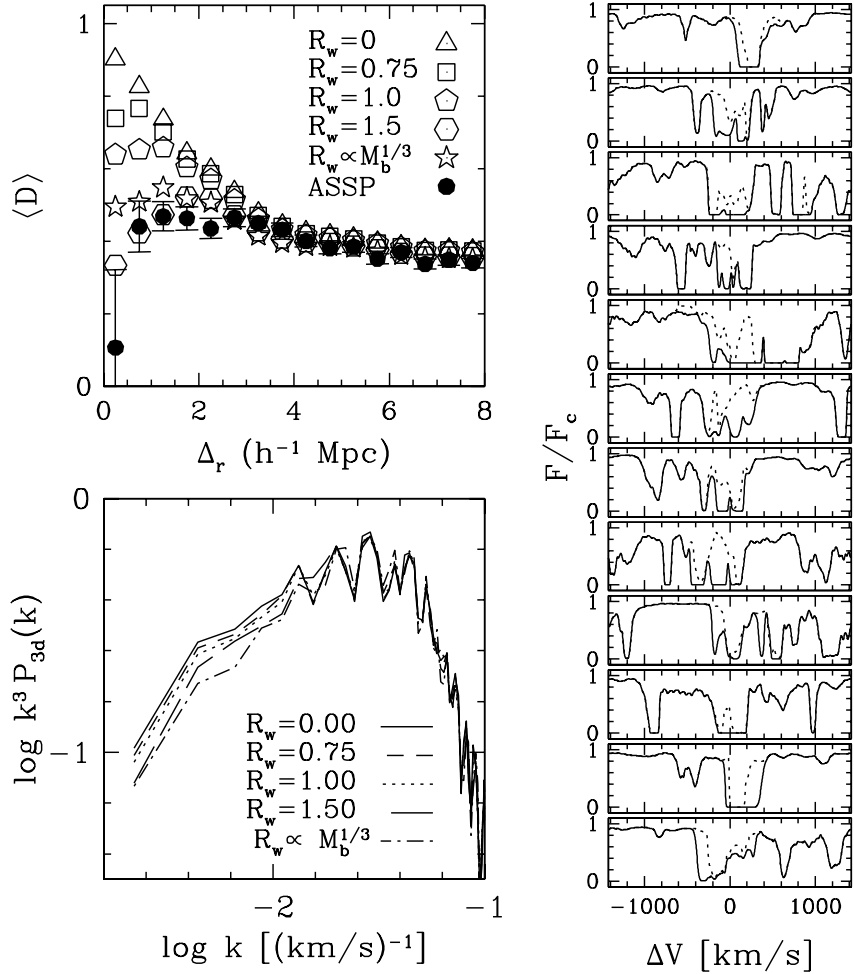}
}
\caption{
Influence of idealized ``winds'' on the conditional mean flux decrement
and the flux power spectrum, based on an SPH simulation of a $22.222\hmpc$
(comoving) cube.  Triangles in the upper left panel show the mean flux 
decrement in pixels that lie at redshift-space separation $\Delta_r$ from 
one of the 40 brightest galaxies in the cube.  Results are averaged over
$0.5\hmpc$ (comoving) bins of $\Delta_r$.  Squares, pentagons, and
hexagons show the effect of removing all neutral hydrogen in spheres
of comoving radius 0.75, 1.0, and $1.5\hmpc$ around these galaxies 
before extracting spectra, and stars show a model with 
sphere volume proportional to galaxy baryon mass.
Filled circles show
the observational estimates of ASSP.  Sample spectra on the right illustrate
the effect qualitatively.  Solid curves show spectra along 12 lines of
sight selected to pass within $0.5\hmpc$ ($24"$) of a target galaxy
at redshift space position $\Delta V=0$.
Dotted curves show the corresponding spectra for $R_w=1.5\hmpc$.
The lower left panel shows the flux power
spectrum for the unmodified SPH simulation (solid) and the various
``wind'' models (as marked).
}
\label{fig:wind}
\end{figure}

Figure~\ref{fig:wind} illustrates this conflict.  In the upper left panel,
triangles show the predicted mean decrement in bins of $\Delta_r$, while
filled circles show the ASSP data points.  The predictions come from an SPH
simulation of a $22.222\hmpc$ cube, and we select the 40 galaxies with
the highest star formation rates to approximate the magnitude limit
of ASSP's spectroscopic survey (see \cite{kollmeier02} for details).
Plausible random errors in galaxy redshifts can reduce the discrepancy
at $\Delta_r\approx 1-2\hmpc$, but they cannot explain the results at
the smallest scales, especially the innermost data point at 
$\Delta_r=0-0.5\hmpc$, $\meand = 0.11$ 
\cite{croft02b,kollmeier02,kollmeier02b}.
The right column of Figure~\ref{fig:wind} shows simulated spectra along
12 lines of sight that pass within $0.5\hmpc$ of a target galaxy, centered
on the galaxy redshift space position.  All but one of these spectra have
$D\geq 0.6$ at the galaxy redshift, and half have $D\geq 0.9$.

\begin{figure}
\centerline{
\epsfxsize=4.5truein
\epsfbox[60 210 555 720]{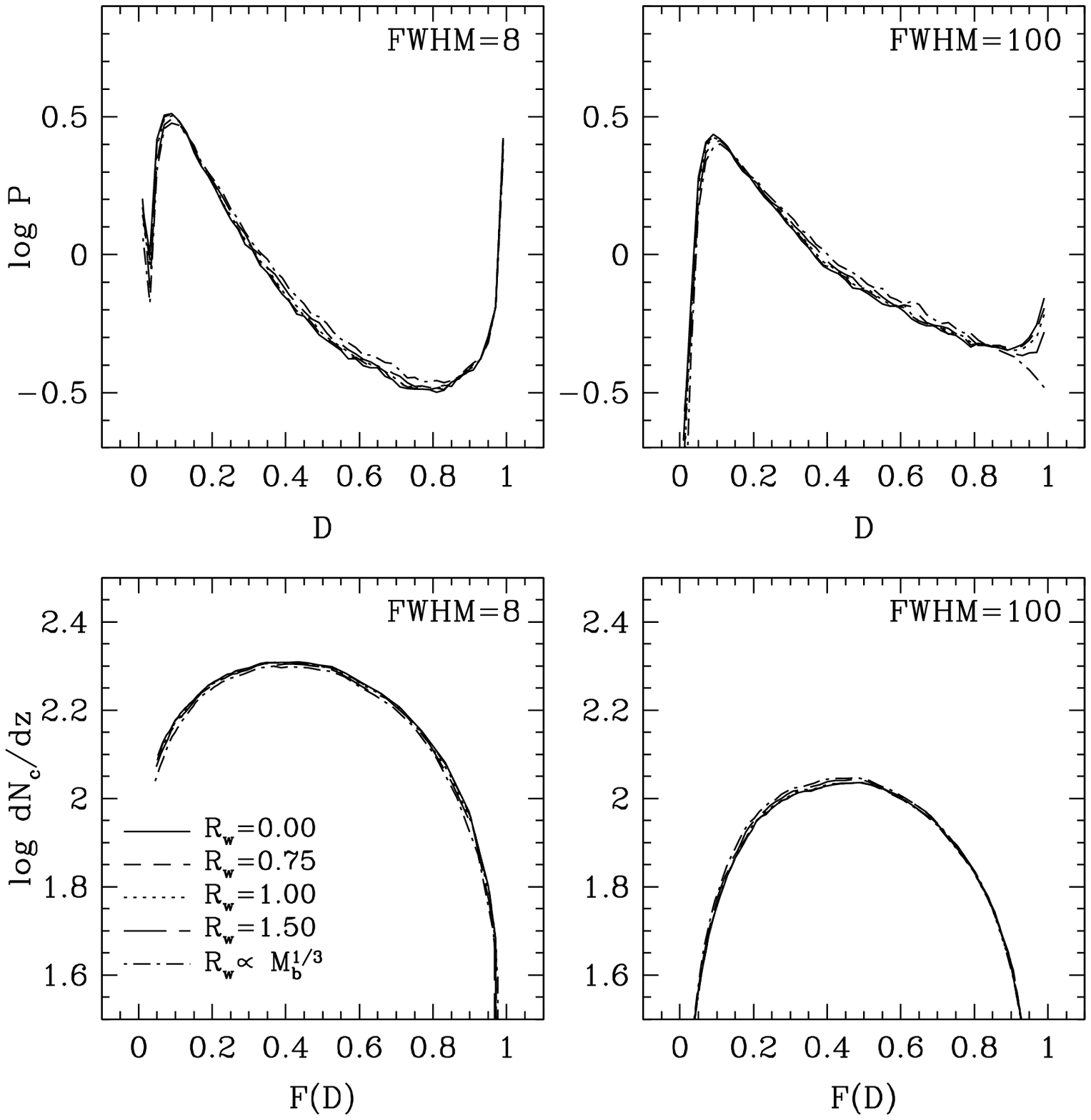}
}
\caption{
Influence of winds on the flux decrement PDF (top, as in Fig.~\ref{fig:fdf})
and the threshold crossing frequency (bottom, as in Fig.~\ref{fig:dcross}).
Results are shown for the $22.222\hmpc$ SPH simulation with no modification
(solid) and with the ``wind'' models illustrated in Fig.~\ref{fig:wind}.
}
\label{fig:windstats}
\end{figure}

The obvious conclusion is that feedback from the observed galaxies 
reduces neutral hydrogen in their immediate surroundings.  Local 
photoionization by the galaxies' stars or active nuclei proves
insufficient; this argument can be cast in general terms that
seem difficult to escape \cite{kollmeier02}.  The natural alternative
is some form of supernova or AGN-driven wind 
\cite{adelberger02,croft02b,bruscoli02}.  Here (and in \cite{kollmeier02b})
we have investigated highly idealized ``wind'' models in which we simply
eliminate all neutral hydrogen in a sphere of comoving radius $R_w$
around each target galaxy.  Squares, pentagons, and hexagons in
Figure~\ref{fig:wind} show results for $R_w=0.75$, 1.0, and $1.5\hmpc$.
Stars show a model where we place winds around all 641 resolved galaxies
in the simulation volume (instead of the top 40 that constitute the 
``observed'' sample) and scale the wind volume in proportion to the
galaxy baryon mass, with a normalization $R_w=1\hmpc$ around the 40th-ranked
galaxy.  In the right column, dotted curves show spectra for the 
$R_w=1.5\hmpc$ model.  
 
The most important lesson from Figure~\ref{fig:wind} is that eliminating
neutral gas to a {\it real space} distance $R_w$ does not eliminate
absorption within a {\it redshift space} separation $R_w$ 
(e.g., to $\Delta V = \pm 200\kms$ in the spectrum plots).
Peculiar infall velocities allow gas at larger distances to produce
absorption near the galaxy redshift, making the energetic requirements
on any wind explanation of the observed \lya\ deficit much more stringent.
Only the mass-scaled and $R_w=1.5\hmpc$ models come close to matching
the ASSP data.  Winds that fully ionize or perfectly entrain gas to
such large distances are not a natural outcome of hydrodynamic 
simulations with stellar feedback \cite{theuns02b,bruscoli02},
and even reaching $1.5\hmpc$ in $\sim 1\;$Gyr
requires a sustained propagation speed $\sim 600\kms$.

Given the challenges facing the wind explanation, it is worth considering
the alternative possibility that extended \lya\ {\it emission} from the 
target galaxies is ``filling in'' the corresponding region of the 
absorption spectrum.  Steidel et al.\ \cite{steidel00} observed two
extended \lya\ ``blobs'' apparently associated with LBGs, with angular
extents $\sim 15"$ and AB apparent magnitudes 21.02 and 21.14 in the \lya\ 
band.  Cooling radiation from gas settling into massive galaxies at $z=3$
naturally produces \lya\ fluxes of this order \cite{haiman00,fardal01}.
The three quasars that contribute to ASSP's innermost data point have
$G$-band AB magnitudes of 20.1, 21.6, and 23.4, so if {\it all} of a
target galaxy's \lya\ cooling radiation went down the slit it could
potentially replace the quasar flux absorbed by the surrounding IGM.
However, a $1.4"$ slit at an angular separation $\Delta\theta \sim 15-20"$
from a galaxy should intercept at most $\sim 1.4/2\Delta\theta \sim 0.03-0.05$
of the galaxy's extended \lya\ flux, at least on average, so this 
explanation seems to fail by $1-2$ orders of magnitude.  Furthermore,
a fourth pair involving the $G=17.8$
quasar Q0302-0019, shows no sign of absorption near the galaxy redshift,
and in this case the quasar is clearly too bright for galaxy emission
to compete with it.  
(This pair and two others are dropped from ASSP's $\meand$ calculation
because of possible Ly$\beta$ contamination.)
At this point, the \lya\ emission explanation 
seems unlikely; it can be conclusively ruled out by
observing more close pairs involving bright quasars or by obtaining
symmetrically placed spectra away from the observed quasars to 
search for galaxy emission.

Assuming for now that winds are the correct explanation for the observed \lya\ 
deficit, one can ask whether they completely spoil the picture
painted in \S\S 1 and 2.  The lower left panel of Figure~\ref{fig:wind}
shows the flux power spectrum for the various ``wind'' models.
Eliminating neutral hydrogen to distances $R_w=0.75$ or $1\hmpc$ around
bright galaxies has only a small impact on the flux power spectrum,
because the filling factor of the ``bubbles'' is small and absorption
close to the galaxies remains nearly saturated in any case.  However, 
these models also do not explain the ASSP results.  In the two more 
extreme models, winds suppress the flux power spectrum on large scales
by $0.1-0.2$ dex, comparable to the 0.1-dex $1\sigma$ uncertainty
that \cite{croft02} quote for the normalization of the
matter power spectrum.  Thus, winds of this magnitude could have 
systematic effects on $P_m(k)$ determinations that are significant
relative to the present observational uncertainties.
The influence of winds on the flux PDF and threshold crossing frequency
is smaller, as shown in Figure~\ref{fig:windstats}.  Even the more
extreme models have a negligible impact on $dN_c/dz$, and they only
slightly alter the shape of the PDF.  The most significant effect is on
the fraction of saturated pixels for $100\kms$ smoothing, which drops
from 2.5\% for the no-wind case to 1.9\% for $R_w=1.5\hmpc$ and 1.3\% 
for the mass-scaled model.

\section{Prospects and Challenges}

From the above discussion, it is clear that one immediate challenge is
to better understand the galaxy proximity effect.  
Figures~\ref{fig:wind} and~\ref{fig:windstats} show that winds can reach
substantial distances from bright galaxies without having much impact
on the global statistics of the \lya\ forest (see also 
\cite{theuns02b,bruscoli02}), but even the more extreme
wind models considered in \S 3 do not reproduce the ASSP results very well.
Simulations with more realistic wind physics 
\cite{croft02b,theuns02b,bruscoli02} can shed further light on this problem,
but major progress will have to await further observational studies,
since current inferences rest crucially on a handful of galaxy-quasar pairs.

Studies with Keck HIRES and VLT UVES have provided a (still growing) trove
of high quality data on the \lya\ forest at $z\sim 2-4$.  The fluctuating
IGM perspective, furthermore, shows that large samples of moderate resolution
spectra can be a powerful resource for studying structure on large scales,
since continuous flux statistics do not require resolution of individual
``lines.''  Such samples can be assembled quickly with large telescopes, 
and the SDSS is producing an enormous sample at resolution $R\sim 2000$ 
in the course of its normal operations.  
One of the observational frontiers is the use of correlations across
multiple lines of sight.  Quasar pair studies provided the first decisive
evidence for large coherence scales of absorbing structures
\cite{bechtold94,dinshaw94,dinshaw95,crotts98}, but larger samples allow
more ambitious goals, such as using the Alcock-Pacyznski test \cite{alcock79}
to measure spacetime geometry \cite{hui99,mcdonald99,mcdonald01}, 
improving measurements of the flux power spectrum with cross correlations
\cite{viel02b},
and mapping large scale 3-dimensional structure at high redshift
\cite{liske00,rollinde02}.

Analyses of existing data have already led to an important cosmological
conclusion, namely that models with matter clustering similar to that
of ``concordance'' $\Lambda$CDM at $z\sim 3$ are consistent with the
observed \lya\ forest while models with substantially different clustering
amplitudes or suppression of small scale power are not
\cite{rauch97,croft99,theuns99,weinberg99,mcdonald00,narayanan00,
theuns00,meiksin01,zaldarriaga01,croft02,gnedin02}.  
They have also provided constraints on 
the temperature of the diffuse IGM \cite{ricotti00,mcdonald00b,schaye00} and
indications of helium reionization at $z\sim 3.2$ 
\cite{schaye00,theuns02c,theuns02d,bernardi03}.
However, to do justice to the quality
and quantity of data and keep pace with the tightening observational
constraints from other observables, we must play for higher stakes.
It looks possible in principle to achieve precision of $5-10\%$ on
quantities like the matter fluctuation amplitude and $H(z)$, but even without
the potential complications of galaxy feedback, inferences at this level
require more extensive theoretical modeling.  Many effects that are
unimportant at the 25\% level --- differences between approximate methods
and full hydrodynamics, numerical resolution and box size limitations,
spatial fluctuations in IGM temperature, details of continuum determination ---
may become critical at the $5-10\%$ level.
Despite these challenges, the \lya\ forest is the most promising
tool we have for precision cosmological measurements at $z\sim 2-4$.
These measurements might, in the end, simply confirm the cosmological
model favored by other data, but complementary constraints have the
potential to break parameter degeneracies and thereby
reveal subtle quantitative discrepancies.  These in turn could yield
insight into the nature of dark energy, the mechanisms of inflation,
or some other fundamental aspect of our universe.

%%%%%%%%%%%%%%%%%%%%%%%%%%%%%%%%%%%%%%%%%%%%%%%%
%% BACKMATTER
%%%%%%%%%%%%%%%%%%%%%%%%%%%%%%%%%%%%%%%%%%%%%%%%

\bibliographystyle{aipproc}

\begin{thebibliography}
\expandafter\ifx\csname natexlab\endcsname\relax\def\natexlab#1{#1}\fi
\providecommand{\enquote}[1]{``#1''}
\expandafter\ifx\csname url\endcsname\relax
  \def\url#1{\texttt{#1}}\fi
\expandafter\ifx\csname urlprefix\endcsname\relax\def\urlprefix{URL }\fi

\bibitem[Bechtold et al.(1994)]{bechtold94}  
Bechtold, J., Crotts, A. P. S., Duncan, R. C., \& Fang, Y. 1994, \apj, 437, L83
% paired lines of sight

\bibitem[Dinshaw et al.(1994)]{dinshaw94}
Dinshaw, N., Impey, C. D., Foltz, C. B., Weymann, R. J., \&
Chaffee, F. H. 1994, \apj, 437, L87
% paired lines of sight

\bibitem[Dinshaw et al.(1995)]{dinshaw95}
Dinshaw, N., Foltz, C. B., Impey, C. D., Weymann, R. J., \&
Morris, S. L. 1995, Nature, 373, 223
% paired lines of sight

\bibitem[Crotts \& Fang(1998)]{crotts98}
Crotts, A. P. S., \& Fang, Y. 1998, \apj, 502, 16
% quasar pairs, triplet

\bibitem[Hu et al.(1995)]{hu95} 
Hu, E.M., Kim, T.S., Cowie, L.L., Songaila, A., \& Rauch, M. 1995, \aj,
110, 1526 
% f(N) and n(b) from Keck spectra

\bibitem[Songaila \& Cowie(1996)]{songaila96} 
Songaila, A. \& Cowie, L.L. 1996, \aj, 112, 335
% metal lines in forest

\bibitem[Ellison et al.(1999)]{ellison99}
Ellison, S.~L., Lewis,
G.~F., Pettini, M., Chaffee, F.~H., \& Irwin, M.~J.\ 1999, \apj, 520, 456
% metals in forest

\bibitem[Bi(1993)]{bi93}   
Bi, H.G., 1993, \apj, 405, 479 
% linear theory model of Lya forest

\bibitem[Cen et al.(1994)]{cen94}  
Cen, R., Miralda-Escud\'e, J., Ostriker, J.P., \& Rauch, M. 1994, \apj, 437, L9
% Ly-alpha forest from gravitational collapse

\bibitem[Petitjean, M\"ucket, \& Kates(1995)]{petitjean95}
Petitjean, P., M\"ucket, J. P., \& Kates, R. E. 1995, A\&A, 295, L9
% low-z ly-a forest, filaments

\bibitem[Zhang, Anninos, \& Norman(1995)]{zhang95}
Zhang, Y., Anninos, P., \& Norman, M.L. 1995, \apj, 453, L57 
% first Illinois ly-a forest paper

\bibitem[Hernquist et al.(1996)]{hernquist96}  
Hernquist L., Katz, N., Weinberg, D.H., \&
Miralda-Escud\'e, J. 1996, \apj, 457, L51
% Lya forest in CDM

\bibitem[Bi \& Davidsen(1997)]{bi97}  
Bi, H.G., \& Davidsen, A. 1997, \apj , 479, 523
% lognormal model of Lya forest

\bibitem[Hui, Gnedin, \& Zhang(1997)]{hui97}
Hui, L., Gnedin, N., \& Zhang, Y. 1997, \apj, 486, 599
% density peak approximation and f(N)

\bibitem[Theuns et al.(1998)]{theuns98}
Theuns, T., Leonard, A., Efstathiou, G., Pearce, F. R., \& Thomas, P. A. 1998,
\mnras, 301, 478
% P3MSPH simulations of lya forest

\bibitem[Gunn \& Peterson(1965)]{gunn65} 
Gunn, J.E., \& Peterson, B.A. 1965, \apj, 142, 1633

\bibitem[Croft et al.(1997)]{croft97} 
Croft, R.A.C., Weinberg, D.H., Katz, N., Hernquist, L., 1997, \apj, 488, 532
% HeII

\bibitem[Rauch et al.(1997)]{rauch97}
Rauch, M., Miralda-Escud\'e, J., Sargent, W. L. W., Barlow, T. A.,
Weinberg, D. H., Hernquist, L., Katz, N., Cen, R., \& Ostriker, J. P.,
1997, \apj, 489, 7
% Omega_b from flux decrement distribution, HIRES vs. simulations

\bibitem[Weinberg, Katz, \& Hernquist(1998)]{weinberg98}
Weinberg, D. H., Katz, N., \& Hernquist, L. 1998,
in ASP Conference Series 148, Origins,
eds. C. E. Woodward, J. M. Shull, \& H. Thronson,
(ASP: San Francisco), 21, astro-ph/9708213

\bibitem[Hui \& Gnedin(1997)]{huig97}
Hui, L., \& Gnedin, N. 1997, \mnras, 292, 27
% IGM equation of state

\bibitem[Weinberg, Hernquist, \& Katz(1997)]{weinberg97}
Weinberg, D. H., Hernquist, L., \& Katz, N. 1997, \apj, 477, 8

\bibitem[Croft et al.(1998)]{croft98}
Croft, R. A. C., Weinberg, D. H., Katz, N., \& Hernquist, L. 1998,
\apj, 495, 44
% P(k)

\bibitem[Weinberg et al.(1999)]{weinberg99}
Weinberg, D. H., et al.\ 1999, in Evolution of Large Scale Structure:
From Recombination to Garching, eds. A.J. Banday, R. K. Sheth,
\& L. N. Da Costa, (Twin Press: Vledder NL), 346 astro-ph/9810142
% Cosmology with the Lyman-alpha Forest

\bibitem[Bi, Ge, \& Fang(1995)]{bi95}
Bi, H., Ge, J., \& Fang, L.-Z. 1995, \apj, 452, 90
% lognormal model of Lya forest

\bibitem[Gnedin \& Hui(1998)]{gnedin98}
Gnedin, N. Y., \& Hui, L. 1998, \mnras, 296, 44
% HPM technique

\bibitem[Zaldarriaga, Hui, \& Tegmark 2001]{zaldarriaga01}
Zaldarriaga, M., Hui, L., \& Tegmark, M.\ 2001, \apj, 557, 519
% constraints from matching 1-d Lya flux P(k)

\bibitem[Viel et al.(2002)]{viel02}
Viel, M., Matarrese, S.,
Mo, H.~J., Theuns, T., \& Haehnelt, M.~G.\ 2002, \mnras, 336, 685 
% modeling IGM from dark matter distribution

\bibitem[Weinberg et al.(1997)]{weinberg97b}
Weinberg, D.H., Miralda-Escud\'{e}, J., Hernquist, L., \& Katz, N., 1997,
\apj, 490, 564
% lower bound on Omega_b

\bibitem[Croft et al.(2002)]{croft02}
Croft, R. A. C., Weinberg, D. H., Bolte, M., Burles, S.,  Hernquist, L., Katz,
N., Kirkman, D., Tytler, D. 2002, \apj, 581, 20
% P(k) measurement

\bibitem[Gnedin \& Hamilton(2002)]{gnedin02}
Gnedin, N.~Y.~\& Hamilton, A.~J.~S.\ 2002, \mnras, 334, 107
% Lya P(k) myth or reality?

\bibitem[Tegmark \& Zaldarriaga(2002)]{tegmark02}
Tegmark, M. \& Zaldarriaga, M. 2002, PRD, 66, 103508
% parameters beyond black box, Lya forest vs. other data (astro-ph/0207047)

\bibitem[McDonald et al.(2000)]{mcdonald00} 
McDonald, P., Miralda-Escud\'e, J., Rauch, M., Sargent, W. L. W.,
Barlow, T. A., Cen, R., \& Ostriker, J. P. 2000, \apj, 543, 1
% PDF, power spectrum, and correlation function of Lya forest flux

\bibitem[Cen(1997)]{cen97}
Cen, R.\ 1997, \apj, 479, L85 
% high end of flux PDF as test of fluctuation amplitude

\bibitem[Miralda-Escud\'e et al.(1996)]{miralda96} 
Miralda-Escud\'e J., Cen R., Ostriker, J.P., \& Rauch, M. 1996, \apj, 471, 582
% Lyman-alpha forest tome

\bibitem[Kujat et al.(2002)]{kujat02}
Kujat, J., Linn, A. M., Scherrer, R. J., \& Weinberg, D. H.\ 2002,
\apj, 572, 1
% prospects for determining equation of state

\bibitem[Viel et al.(2002)]{viel02c}
Viel, M., Matarrese, S., Theuns, T., Munshi, D., \& Wang, Y.\ 2002,
\mnras, submitted, astro-ph/0212241
% dark energy effects on lya forest

\bibitem[McDonald, Miralda-Escud{\' e}, \& Cen(2002)]{mcdonald02}
McDonald, P., Miralda-Escud{\' e}, J., \& Cen, R.\ 2002, \apj, 580, 42
% mass-lya correlations

\bibitem[Croft et al.(2002)]{croft02b}
Croft, R. A. C., Hernquist, L., Springel, V., Westover, M., White, M.\ 2002,
\apj, 580, 634
% LBGs and Lya forest 

\bibitem[Kollmeier et al.(2002)]{kollmeier02}
Kollmeier, J.A., Weinberg, D.H., Dav\'{e}, R., Katz, N., 2002, ApJ, submitted,
astro-ph/0209563 
%LBG-Forest correlations

\bibitem[Bruscoli et al.(2002)]{bruscoli02}
Bruscoli, M., Ferrara, A., Marri, S., Schneider, R., Maselli, A.,
Rollinde, E., Aracil, B., \mnras, submitted, astro-ph/0212126
%LBG-Forest correlations-Winds 

\bibitem[Adelberger et al.(2002)]{adelberger02}
Adelberger, K.L., Steidel, C.C., Shapley, A.E., Pettini, M. 2002,
ApJ, in press, astro-ph/0210314
% Galaxy Proximity Effect

\bibitem[Kollmeier et al.(2002b)]{kollmeier02b}
Kollmeier, J.A., Weinberg, D.H., Dav\'{e}, R., \& Katz, N., these proceedings

\bibitem[Theuns et al.(2002)]{theuns02b}
Theuns, T., Viel, M., 
Kay, S., Schaye, J., Carswell, R.~F., \& Tzanavaris, P.\ 2002, \apj, 578,
L5
% winds and IGM: winds produce enrichment, suppress galaxies, w/o changing Lyaf

\bibitem[Steidel et al.(2000)]{steidel00}
Steidel, C. C., Adelberger, K. L., Shapley, A. E., Pettini, M.,
Dickinson, M., \& Giavalisco, M. 2000, \apj, 532, 170
% Lya imaging of z=3 proto-cluster, Lya blobs

\bibitem[Haiman, Spaans, \& Quataert(2000)]{haiman00}
Haiman, Z., Spaans, M., \& Quataert, E. 2000, \apj, 537, L5
% lya cooling radiation from high-z halos

\bibitem[Fardal et al.\ (2001)]{fardal01}
Fardal, M.\ A., Katz, N., Gardner, J.\ P., Hernquist, L.,
Weinberg, D.\ H.\ \& Dav{\'e}, R.\ 2001, \apj, 562, 605
% cooling radiation and lya luminosity

\bibitem[Alcock \& Paczy\'nski(1979)]{alcock79}
Alcock, C., \& Paczy\'{n}ski, B. 1979, Nature, 281, 358
% distortions and lambda

\bibitem[Hui, Stebbins, \& Burles(1999)]{hui99}
Hui, L., Stebbins, A., \& Burles, S. 1999, \apj, 511, 5
% Geometrical test using Lya forest pairs

\bibitem[McDonald \& Miralda-Escud\'e(1999)]{mcdonald99}
McDonald, P. \& Miralda-Escud\'e, J. 1999, \apj, 518, 24
% Geometry from Lya forest pairs

\bibitem[McDonald(2001)]{mcdonald01}
McDonald, P. 2001, \apj, submitted, astro-ph/0108064
% lya P(k) in 3-d, towards measurement of geometry

\bibitem[Viel et al.(2002)]{viel02b}
Viel, M., Matarrese, S., 
Mo, H.~J., Haehnelt, M.~G., \& Theuns, T.\ 2002, \mnras, 329, 848
% lya absorption along multiple lines of sight, recovery of P(k)

\bibitem[Liske et al.(2000)]{liske00}
Liske, J., Webb, J.~K.,
Williger, G.~M., Fern{\' a}ndez-Soto, A., \& Carswell, R.~F.\ 2000, \mnras, 
311, 657 
% lya absorption towards group of 10 quasars

\bibitem[Rollinde et al.(2002)]{rollinde02}
Rollinde, E., Petitjean, P., Pichon, C., Colombi, S., Aracil, B.,
D'Odorico, V., \& Haehnelt, M. G. 2002, \mnras, submitted
% lya forest correlation in close pairs and groups

\bibitem[Croft et al.(1999))]{croft99}
Croft, R. A. C., Weinberg, D. H., Pettini, M., Katz, N., \& Hernquist, L. 1999,
\apj, 520, 1
% P(k) measurement

\bibitem[Theuns, Leonard, Schaye, \& Efstathiou(1999)]{theuns99}
Theuns, T., Leonard, A., Schaye, J., \& Efstathiou, G.\ 1999, \mnras, 303, L58
% line statistics comparison to observe Lyaf, high-T required

\bibitem[Narayanan et al.(2000)]{narayanan00}
Narayanan, V.~K., Spergel, D.~N., Dav{\' e}, R., \& Ma, C.\ 2000, 
\apj, 543, L103 
% Lyaf constraints on warm dark matter

\bibitem[Theuns, Schaye, \& Haehnelt(2000)]{theuns00}
Theuns, T., Schaye, J., \& Haehnelt, M.~G.\ 2000, \mnras, 315, 600 
% constraining temperature with P(k), agreement of FDF w/ LCDM, T=15000

\bibitem[Meiksin, Bryan, \& Machacek(2001)]{meiksin01}
Meiksin, A., Bryan, G., \& Machacek, M.\ 2001, \mnras, 327, 296 
% comparison of simulations and observations of lyaf

\bibitem[Ricotti, Gnedin, \& Shull(2000)]{ricotti00}
Ricotti, M., Gnedin, N.\ Y.\ \& Shull, J.\ M.\ 2000, \apj, 534, 41 
% evolution of IGM rho-T relation

\bibitem[McDonald et al.(2000)]{mcdonald00b}
McDonald, P., Miralda-Escud\'e, J., Rauch, M., Sargent, W. L. W.,
Barlow, T. A., \& Cen, R. 2000, \apj, 562, 52
% temperature-density relation

\bibitem[Schaye et al.(2000)]{schaye00}
Schaye, J., Theuns, T., 
Rauch, M., Efstathiou, G.\ \& Sargent, W.\ L.\ W.\ 2000, \mnras, 318, 817 
% measurement of IGM equation of state

\bibitem[Theuns et al.(2002)]{theuns02c}
Theuns, T., Zaroubi, S.,
Kim, T., Tzanavaris, P., \& Carswell, R.~F.\ 2002, \mnras, 332, 367
% wavelet evidence for HeII reionization at z~3.3

\bibitem[Theuns et al.(2002)]{theuns02d}
Theuns, T., Bernardi, M., Frieman, J., Hewett, P., Schaye, J.,
Sheth, R.~K., \& Subbarao, M.\
2002, \apj, 574, L111
% HeII reionization in SDSS spectra

\bibitem[Bernardi et al.(2003)]{bernardi03}
Bernardi, M., et al.\ 2003, \aj, 125, 32
% feature in optical depth at z~3.2

\end{thebibliography}

\end{document}